\documentclass{PoS}

\title{Measurement of the transverse target-spin asymmetry associated with deeply virtual Compton scattering on the proton}

\ShortTitle{Measurement of the TTSA associated with DVCS on the proton}

\author{\speaker{Zhenyu Ye}\thanks{on behalf of the HERMES collaboration}\\

        DESY, 22607 Hamburg, Germany\\

        E-mail: \email{yezhenyu@mail.desy.de}}

\abstract{Measurements of deeply virtual Compton scattering (DVCS)
made at HERMES using 27.6 GeV $e^\pm$ beams and various internal
polarized or unpolarized gaseous targets are discussed. Results
are reported on the transverse target-spin asymmetry (TTSA)
associated with DVCS, extracted from data accumulated in 2002-2004
with the $e^+$ beam on a transversely polarized hydrogen target.
TTSA amplitudes leading in twist and $\alpha_S$ are given as a
function of $-t$, $x_B$ and $Q^2$ in the kinematic range
$\left|t\right|<0.7$ GeV$^2$, $0.03<x_B<0.35$ and $1<Q^2<10$
GeV$^2$. Theoretical predictions based on a phenomenological model
of generalized parton distributions (GPDs) agree with the
experimental results. With additional statistics accumulated in 2005,
one may constrain the $u$-quark total angular momentum in the nucleon
within this model.}

\FullConference{International Europhysics Conference on High Energy Physics\\
         July 21st - 27th 2005\\
         Lisboa, Portugal}

\begin{document}

\section{Introduction}
Deeply virtual Compton scattering (DVCS) is an exclusive process
in which a virtual photon (emitted by an incoming lepton) is absorbed
and a real photon is produced by a single parton in the nucleon, the
recoiling nucleon being in its ground state. It is one of the theoretically
cleanest processes to access generalized parton distributions (GPDs);
the GPDs provide a detailed description of the nucleon
structure. Strong interest in DVCS arose recently after it was
realized that GPDs encode information about the total angular
momentum of the partons in the nucleon \cite{Ji97}.

\begin{figure}[b]\begin{center}
\includegraphics[angle=270,width=0.45\columnwidth]{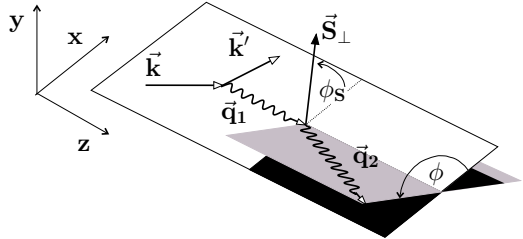}
\caption{Kinematics and azimuthal angles of real photon
electroproduction in the target rest frame. The $z$-direction is
chosen along the three-momentum of the virtual photon $\vec{q}_1$.
The lepton three-momenta $\vec{k}$ and $\vec{k}'$ form the lepton
scattering plane, while the three-momenta of virtual and real
photons $\vec{q}_1$ and $\vec{q}_2$ define the photon production
plane. The azimuthal angles $\phi$ and $\phi_S$ are explained in the
text.\label{fig:coordinate}}
\end{center}\end{figure}

The final state of DVCS is identical to that of the Bethe-Heitler (BH)
process, in the latter the real photon being radiated from the
incoming or scattered lepton. Hence the experimental cross section
contains the interference between the BH amplitude
$\mathcal{T}_{BH}$ and the DVCS amplitude $\mathcal{T}_{DVCS}$:
\begin{equation}
\frac{d\sigma}{dx_B\,dQ^2\,d\left|t\right|\,d\phi\,d\phi_S} \propto
\left|\mathcal{T}_{BH}\right|^2+\left|\mathcal{T}_{DVCS}\right|^2+\mathcal{T}_{BH}^*\mathcal{T}_{DVCS}+\mathcal{T}_{BH}\mathcal{T}_{DVCS}^*.
\end{equation}
Here $x_B$ denotes the Bjorken scaling variable, $-Q^2$ the square
of the 4-momentum of the virtual photon and $t$ the square of the
4-momentum transfer between initial and final target nucleons. The
angle $\phi$ ($\phi_S$) denotes the azimuthal angle of the photon
production plane (the target polarization vector) with respect to
the lepton scattering plane, measured around the direction of the
virtual photon (see figure \ref{fig:coordinate}).

While $\mathcal{T}_{DVCS}$ contains information about unknown
GPDs, $\mathcal{T}_{BH}$ is exactly calculable in QED using the
knowledge of the well-determined nucleon form factors
\cite{belitsky}. In fixed-target experiments at moderate $Q^2$
(e.g. at HERMES \cite{HERMES}), where the BH contribution widely dominates
the cross section, DVCS results have already been obtained by measuring
the azimuthal asymmetry in the cross section with respect to the beam
helicity \cite{HERMES-BSA}, to the beam charge \cite{HERMES-BCA},
and to the longitudinal target-spin orientation \cite{HERMES-LTSA}. In this
article we will report the first results on the transverse
target-spin asymmetry (TTSA) obtained at HERMES.

\section{The HERMES Experiment}
HERMES studies the spin structure of the nucleon using the 27.6
GeV electron (or positron) beam at HERA and internal polarized H, D,
and $^3$He gaseous targets. A spectrometer magnet instrumented with
tracking chambers provides good momentum ($\delta p/p<2\%$) and angular
($\delta \theta<1$ mrad) measurements for charged particles \cite{HERMES}.
Good lepton-hadron separation is achieved by a transition-radiation
detector, a pre-shower counter and an electromagnetic calorimeter,
which also detects the photon.

Not all the hadrons in the final state of the BH and DVCS
processes are detected at HERMES by its forward spectrometer --
the recoiling proton typically travels perpendicularly to the beam
direction and hence escapes the detector acceptance. The
exclusivity of the selected events, which contain an identified
scattered lepton and a produced real photon, is maintained by a
missing-mass cut of $-(1.5)^2<M_x^2<(1.7)^2$ GeV$^2$. Monte Carlo
studies have shown that the non-DVCS contributions to the
selected data sample originates mainly from semi-inclusive $\pi^0$
production amounting to approximately $5\%$ \cite{Krauss05}.

\section{Transverse Target-Spin Asymmetry}
The transverse target-spin asymmetry, measured using an
unpolarized (U) lepton beam and a transversely (T) polarized
target, is defined as:
\begin{equation}
A_{UT}(\phi,\phi_S)=\frac
{d\sigma(\phi,\phi_S) - d\sigma(\phi,\phi_S+\pi)}
{d\sigma(\phi,\phi_S) + d\sigma(\phi,\phi_S+\pi)}. \label{eqn:tta}
\end{equation}
The two azimuthal amplitudes
$A_{UT}^{\sin(\phi-\phi_S)\cos{\phi}}$ and
$A_{UT}^{\cos(\phi-\phi_S)\sin{\phi}}$, which are associated with DVCS
on the proton, can be approximated in leading twist and $\alpha_S$
as \cite{Elli05}:
\begin{eqnarray}
A_{UT}^{\sin{(\phi-\phi_S)}\cos{\phi}} &\propto& \pm f(x_B,Q^2,t) \cdot Im \left[F_2\mathcal{H}-F_1\mathcal{E}\right], \nonumber \\
A_{UT}^{\cos{(\phi-\phi_S)}\sin{\phi}} &\propto& \pm f(x_B,Q^2,t) \cdot Im \left[F_2\mathcal{\widetilde{H}}-F_1\xi\mathcal{\widetilde{E}}\right].\label{eqn:ref1}
\end{eqnarray}
Here $+$ ($-$) stands for a negatively (positively) charged lepton
beam and $f(x_B,Q^2,t)$ denotes a kinematic factor that is
independent on the azimuthal angles. ${\mathcal{H}}$,
${\mathcal{E}}$, ${\mathcal{\widetilde{H}}}$ and
${\mathcal{\widetilde{E}}}$ denote Compton form factors which are
convolutions of the respective twist-2 GPDs $H$, $E$,
$\widetilde{H}$ and $\widetilde{E}$ \cite{belitsky}. $F_1$ and $F_2$
are the Dirac and Pauli form factors of the proton, respectively.
Note that dependences on kinematic variables are partially omitted in equation
(\ref{eqn:ref1}) for clarity.

\begin{figure}[t]\begin{center}
\includegraphics[width=0.65\columnwidth]{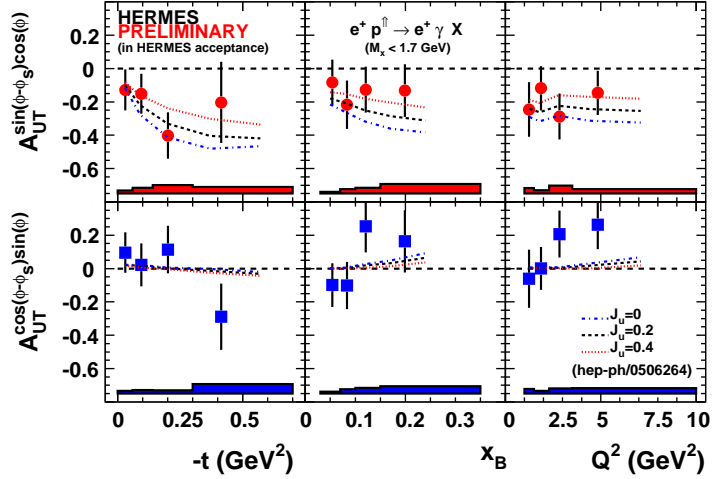}
\caption{The $A_{UT}^{\sin{(\phi-\phi_S)}\cos{\phi}}$ and $A_{UT}^{\cos{(\phi-\phi_S)}\sin{\phi}}$ amplitudes of the transverse target-spin asymmetry associated with DVCS, shown as a function of $-t$, $x_B$ and $Q^2$ for the exclusive sample ($-(1.5)^2<M_x^2<(1.7)^2$ GeV$^2$) after background correction. The error bars (bands) represent the statistical (systematic) uncertainties. The curves are the predictions from a GPD model with different $u$-quark total angular momentum $J_u$ and fixed $d$-quark total angular momentum $J_d=0$ \cite{Elli05}.\label{fig:results}}
\end{center}\end{figure}

Figure \ref{fig:results} shows
$A_{UT}^{\sin{(\phi-\phi_S)}\cos{\phi}}$ and
$A_{UT}^{\cos{(\phi-\phi_S)}\sin{\phi}}$ as a function of $-t$,
$x_B$ and $Q^2$, as extracted at HERMES. The analysis is based on
data accumulated in 2002-2004 with the positron beam and a
transversely polarized hydrogen target. Corrections for background
and smearing have been applied. The main contributions to the
systematic uncertainty are those from the determination of the
target polarization and in the background correction. Also shown
in figure \ref{fig:results} are theoretical predictions based on a
phenomenological model of GPDs \cite{VGG}. The curves represent
the TTSA amplitudes evaluated with different $u$-quark total
angular momentum $J_u$ as a model parameter, while fixing the $d$-quark 
total angular momentum $J_d=0$ \cite{Elli05}. The experimental results 
are in general agreement with the theoretical predictions. More precise 
asymmetries could clearly distinguish between predictions
with different $J_u$, within this model.

\section{Summary and Outlook}
The transverse target-spin asymmetry associated with DVCS on the
proton has been measured at HERMES. An investigation \cite{Elli05} has shown
that a measurement of the TTSA associated with DVCS could lead to constraint
of the $u$-quark total angular momentum in the nucleon
within certain GPD models. Such an analysis is in
progress, which will include the data accumulated at
HERMES in 2005.

\end{document}